\documentclass[preprint,10pt]{elsarticle}

\usepackage{lmodern}
\usepackage[english]{babel}
\usepackage{textcomp} 
\usepackage{amsmath}
\usepackage{mathrsfs}
\usepackage{latexsym}
\usepackage{amssymb}	
\usepackage{amsfonts}
\usepackage{theorem}
\usepackage{graphicx}
\usepackage{xcolor}
\usepackage{setspace}
\usepackage{framed}
\usepackage{hyperref} 
\usepackage{pgf,tikz,pgfplots} 
\usepackage{qrcode} 
\usepackage{multicol}
\usepackage{multirow}
\usepackage{xurl}
\usepackage{tabularx}
\usepackage{enumitem}
\usepackage{float}
\usepackage{colortbl,rotating,booktabs} 
\usepackage{algorithm}
\usepackage{algpseudocode}
\usepackage{caption}

\newcommand{\mb}{\mathbf{m}}
\newcommand{\xb}{\mathbf{x}}
\newcommand{\Xb}{\mathbf{X}}

\newcommand{\Fb}{\mathbf{F}}
\newcommand{\Pb}{\mathbf{P}}
\newcommand{\Qb}{\mathbf{Q}}

\newcommand{\Rb}{\mathbf{R}}

\newcommand{\Zb}{\mathbf{Z}}
\newcommand{\zb}{\mathbf{z}}

\newcommand{\Hb}{\mathbf{H}}
\newcommand{\Sb}{\mathbf{S}}
\newcommand{\Kb}{\mathbf{K}}
\newcommand{\muB}{\mbox{\boldmath$\mu$}}
\journal{}

\begin{document}

\begin{frontmatter}

\title{Chernoff fusion of Bernoulli Gaussian max filters}

\author{Zhijin Chen} 

\affiliation{organization={School of Engineering, RMIT University},
            city={Melbourne},
            postcode={3000}, 
            state={Victoria},
            country={Australia}}
\author{Branko Ristic} 

\author{Du Yong Kim} 


\begin{abstract}
Statistical dependencies between information sources are rarely known, yet in practical  distributed tracking schemes, they must be taken into account in order to prevent track divergences. Chernoff fusion is well-known and universally accepted method that can address the problem of track fusion when the statistical dependence between the fusing sources is unknown. In this paper we derive the exact Chernoff fusion equations for Bernoulli Gaussian max filters. These filters have been recently derived in the framework of possibility theory, as the analog of the Bernoulli Gaussian sum filters. The main motivation for the possibilistic approach is that it effectively deals with imprecise mathematical models (e.g. dynamics, measurements) used in tracking algorithms. The paper also demonstrates the proposed possibilistic fusion scheme in the absence of knowledge about statistical dependence.        

\end{abstract}

\begin{keyword}
Sensor fusion \sep possibility theory \sep Bernoulli filter \sep Gaussian mixture \sep target tracking  


\end{keyword}

\end{frontmatter}



\section{Introduction}
\label{sec1}
Bernoulli filter is the optimal Bayes filter for a single dynamic system which can randomly switch-on or off. It has been first derived for a standard measurement model\footnote{Standard measurement model in target tracking refers to point measurements, affected by clutter and with the probability of detection less than one.} as a JoTT filter in \cite{mahler2007statistical}, and  later popularised in \cite{ristic2013bernoulli}.
In general, Bernoulli filters have no analytic solution and therefore are typically implemented via approximations, such as the particle filter or a Gaussian-sum filter \cite{vo2012multi}, \cite{ristic2013tutorial}. 

Chernoff fusion is well-known and universally accepted method  of track fusion when the statistical dependence between the fusing sources is unknown \cite{julier2006empirical}.
The closed form analytical expressions for Chernoff fusion of Bernoulli Gaussian sum filters were reported in \cite{guldogan2014consensus}. Derivations are based on an approximation  (proposed in 
\cite{julier2006empirical}, see also \cite{gunay2014approximate}) where a weighted Gaussian sum, raised to a power, is approximated by a weighted Gaussian sum. 

 Utilizing the theoretical foundations introduced in \cite{houssineau2018detection}, Ristic et al. \cite{ristic2020target} formulated the Bernoulli filter in the framework of possibility theory. The possibilistic theoretical framework is promoted because of its ability to deal with imperfect (partially known) mathematical models required in target tracking (e.g. measurement models, motion models). More  recently, other tracking filters have been proposed in the possibilistic framework \cite{cai2022possibility, chen2023bernoulli}. Bernoulli Gaussian max filter, as a possibilistic analog of the Bernoulli Gaussian sum filter, was proposed in  \cite{houssineau2022decentralised}. 

 Following the presentation in \cite{guldogan2014consensus}, we formulate the Chernoff fusion algorithms in the framework of possibility theory for Bernoulli Gaussian max filters. Furthermore, we demonstrate the proposed possibilistic fusion scheme in the absence of  knowledge about the statistical dependence.

The paper is organized as follows: Sec. \ref{sec2} will give a brief introduction on Bernoulli Gaussian-max filter (GMF), Sec. \ref{sec3} will formulate Chernoff fusion algorithms of possibilistic Bernoulli filter in both general form and Gaussian-max form. Numerical test results for single target tracking is presented in Sec. \ref{sec4}, and conclusions are drawn in Sec. \ref{sec5}.  

\section{Bernoulli Gaussian max filter}
\label{sec2}

The concept of {\em uncertain variable} in the framework of possibility theory, plays the same role as a random variable in probability theory. The main difference is that the quantities of interest are not random, but simply unknown, and our aim is to infer their true values out of a set of possible values. The theoretical basis of this approach can be found in \cite{bishop2018spatio} and \cite{jeremie_linear_21}. Our knowledge about uncertain  is represented by a possibility function (not a density function), which is noramlised in the sense that its supremum equals 1.

Bernoulli GMF is an approximation of the Bernoulli possibilistic filter, derived in \cite{ristic2020target}, applicable to linear-Gaussian systems.
The posterior possibility function of Gaussian max filter at time $k$ is represented by a triple:
\begin{itemize}
\item $q^0_{k|k}$, the posterior possibility of target absence
\item $q^1_{k|k}$, the posterior possibility of target presence
\item $s_{k|k}(\xb)$, the posterior spatial possibility function, defined over the state space $\mathcal{X}$.
\end{itemize}
Due to normalisation, $\max\{q^0_{k|k},q^1_{k|k}\}=1$ and $\sup_{\xb\in\mathcal{X}} s_{k|k}(\xb) = 1$.

Suppose the target motion model is linear and specified by:
\begin{equation}
\rho_{k|k-1}(\xb|\xb') = \mathcal{\overline{N}}(\xb;\Fb_{k-1}\xb',\Qb_{k-1}) 
\label{e:trans}
\end{equation}
where $\mathcal{\overline{N}}(\xb;\muB,\Pb)$ is the Gaussian possibility function, defined as
\begin{equation}
 \mathcal{\overline{N}}(\xb;\muB,\Pb) = \exp\left\{-\frac{1}{2}(\xb-\muB)^T\;\Pb^{-1}\;(\xb-\muB) \right\}
 \label{e:GPF}
 \end{equation}
for some $\muB\in\mathbb{R}^{n_x}$ and some $n_x\times n_x$ semi-positive matrix $\Pb$. The pair $(\muB,\Pb)$ completely specifies the Gaussian possibility function. With abuse of language, we will refer to $\muB$ and $\Pb$ as to the mean and covariance matrix, respectively.

Under the similar linear Gaussian assumptions, the spatial possibility function $s_{k|k}(\xb)$ can be represented as a Gaussian max mixture, and propagated analytically as a triplet of weights, means and covariances $\{\omega_k^{(i)},\mb_k^{(i)},\Pb_k^{(i)}\}_{i=1}^{\bar{N}_k}$, where $\bar{N}_k$ is the number of possibilistic Gaussian components, $\mb$ is the mean vector and $\Pb$ is the covariance matrix of the Gaussian possibility function $\bar{\mathcal{N}}(\xb;\mb,\Pb)$.
Suppose the spatial possibility function at time $k-1$ is expressed in the Gaussian max mixture form:
\begin{equation}
    s_{k-1|k-1}(\xb)=\displaystyle\max_{1\leq j \leq \bar{N}_{k-1}}\omega_{k-1}^{(j)}\bar{\mathcal{N}}(\xb;\mb_{k-1}^{(j)},\Pb_{k-1}^{(j)}),
\end{equation}
where $\displaystyle\max_{1\leq j \leq \bar{N}_{k-1}}\omega_{k-1}^{(j)}=1$.

In order to model target appearance and disappearance, a binary uncertain variable $\delta_k \in \{0, 1\}$ is introduced, with $\delta_k =1$ indicates that the target is present at time $k$ and $\delta_k =0$ indicates that it is absent. The dynamics of $\delta_k$ is modeled by a two-state Markov chain with a (time-invariant) transitional possibility matrix as follows:
\begin{equation}
    \Phi = \left[\begin{matrix} \tau_{00} & \tau_{01} \\ \tau_{10} &\tau_{11}\end{matrix}\right]
\end{equation}
where $\delta_{ij}$ is the possibility of transition from $\delta_{k-1} =i $ to $\delta_k = j$ for $i,j \in \{0, 1\}$, and $\max\{\delta_{i0},\delta_{i1}\} = 1$, for $i= 0, 1$. The initial possibility (at $k = 0$) of target absence and presence is expressed by total ignorance, i.e.   $\max\{q_0^0,q_0^1\}=1$. The \em possibilistic birth density \em at time $k$, $\bar{b}_{k|k-1}(\xb)$, is also expressed as a Gaussian max mixture: 
\begin{equation}
    \bar{b}_{k|k-1}(\xb)=\displaystyle\max_{1\leq i\leq N_{b,k}}\omega_{b,k}^{(i)}\bar{\mathcal{N}}(\xb;\mb_{b,k}^{(i)},\Pb_{b,k}^{(i)}),
\end{equation}
(where $\displaystyle\max_{1\leq i\leq N_{b,k}}\omega_{b,k}^{(i)}=1$). 

Assume at time $k-1$, the possibility of target existence is $q_{k-1|k-1}^1$ and the possibility of target non-existence is $q_{k-1|k-1}^0$, the prediction stage for the Bernoulli GMF then can be given by the following equations:
\begin{align}
    q_{k|k-1}^0 & =\max\{\tau_{00}q_{k-1|k-1}^0,\tau_{10}q_{k-1|k-1}^1\}, \\
  q_{k|k-1}^1 & =\max\{\tau_{01}q_{k-1|k-1}^0,\tau_{11}q_{k-1|k-1}^1\},\\
  s_{k|k-1}(\xb)&=\frac{1}{q_{k|k-1}^1}\max\{\tau_{01}q_{k-1|k-1}^0\bar{b}_{k|k-1}(\xb),\tau_{11}q_{k-1|k-1}^1 \nonumber \\ &\;\;\; \displaystyle\max_{1\leq i\leq \bar{N}_{k-1}}\omega_{k-1}^{(i)}\bar{\mathcal{N}}(\xb;m_{k|k-1}^{(i)},\Pb_{k|k-1}^{(i)})\},
\end{align}
where
\begin{equation}
    \mb_{k|k-1}^{(i)}=\Fb_{k|k-1}\mb_{k-1}^{(i)},
\end{equation}
\begin{equation}
    \Pb_{k|k-1}^{(i)}=\Qb_{k-1}+\Fb_{k-1}\Pb_{k-1}^{(i)}\Fb_{k-1}^{\intercal},
\end{equation}
The predicted spatial possibility function can also been written in a straightforward Gaussian max form:
\begin{equation}
    s_{k|k-1}^{(i)}(\xb)=\displaystyle\max_{1\leq i \leq \bar{N}_{k|k-1}}\omega_{k|k-1}^{(i)}\bar{\mathcal{N}}(\xb;\mb_{k|k-1}^{(i)},\Pb_{k|k-1}^{(i)}),
\end{equation}
where $\displaystyle\max_{1\leq i \leq N_{k|k-1}}\omega_{k|k-1}^{(i)}=1$.
Then the possibility of absence and possibility of presence can be updated as:
\begin{equation}
    q_{k|k}^0 = \frac{q_{k|k-1}^0}{\max\{q_{k|k-1}^0, \theta q_{k|k-1}^1\}}
\end{equation}
and
\begin{equation}
    q_{k|k}^1 = \frac{\theta q_{k|k-1}^1}{\max\{q_{k|k-1}^0, \theta q_{k|k-1}^1\}},
\end{equation}
respectively, where 
\begin{equation}
    \theta = \max\bigl\{d^0, d^1\displaystyle\max_{\zb\in\Zb_k}\big[\frac{\kappa(\Zb\setminus\{\zb\})}{\kappa(\Zb)}\sup_{\xb\in\mathcal{X}}[g_k(\zb|\xb)\rho_{k|k-1}(\xb)]\big]\bigr\}.
\end{equation}
($d^1$ is the possibility of detection and  $d^0$ is the possibility of non-detection \cite{ristic2020target}).

The spatial possibility function $s_{k|k}(\xb)$ can be updated in the form of a summation of two components, with $s_{k|k}^{ND}(\xb)$ representing no detection cases and $s_{k|k}^{D}(\xb)$ representing the detected target cases.
\begin{equation}
s_{k|k}(\xb) = \max\{s_{k|k}^{ND}(\xb),\displaystyle\max_{\zb\in\Zb_k}s_{k|k}^D(\xb,\zb)\}
\end{equation}
 
For the component of no detection cases, $s_{k|k}^{ND}(\xb)$ is defined in a Gaussian max form:
\begin{equation}
    s_{k|k}^{ND}(\xb)=\displaystyle\max_{1\leq i\leq \bar{N}_{k|k-1}}\omega_{k|k}^{(i)}\mathcal{N}(\xb;\mb_{k|k}^{(i)},\Pb_{k|k}^{(i)}),
\end{equation}
where 
\begin{equation}
    \mb_{k|k}^{(i)}=\mb_{k|k-1}^{(i)},
\end{equation}
\begin{equation}
    \Pb_{k|k}^{(i)}=\Pb_{k|k-1}^{(i)},
\end{equation}
\begin{equation}
    \omega_{k|k}^{(i)}=\frac{d^0}{\theta}\omega_{k|k-1}^{(i)}.
\end{equation}
Similarly, the posterior spatial possibility function $s_{k|k}^D(\xb,\zb)$ which represents the component of the detected cases can be given as:
\begin{equation}
    s_{k|k}^D(\xb,\zb)=\displaystyle\max_{1\leq i\leq \bar{N}_{k|k-1}} \omega_{k|k}^{(i)}\bar{\mathcal{N}}(\xb;\mb_{k|k}^{(i)},\Pb_{k|k}^{(i)}),
\end{equation}
with the propagation: 
\begin{equation}
    \mb_{k|k}^{(i)}=\mb_{k|k-1}^{(i)}+\Tilde{\Kb_k}^{(i)}(\zb-\eta_{k|k-1}^{(i)}),
\end{equation}
\begin{equation}
    \Pb_{k|k}^{(i)}=\Pb_{k|k-1}^{(i)}-\Pb_{k|k-1}^{(i)}\Hb^{\intercal}[\Sb_{k|k-1}^{(i)}]^{-1}\Hb\Pb_{k|k-1}^{(i)},
\end{equation}
\begin{equation}
    \omega_{k|k}^{(i)}=\frac{d^1}{\theta} \displaystyle\max_{\zb\in\Zb_k}\{\frac{\mathcal{N}(\zb; \Hb\mb_{k|k-1}^{(i)}, \Hb\Pb_{k|k-1}^{(i)}\Hb^{\intercal}+R)}{\lambda c(\zb)} \omega_{k|k-1}^{(i)}\},
\end{equation}
where 
\begin{equation}
    \eta_{k|k-1}^{(i)}=\Hb\mb_{k|k-1}^{(i)},
\end{equation}
\begin{equation}
    \Sb_{k|k-1}^{(i)}=\Hb\Pb_{k|k-1}^{(i)}\Hb^{\intercal}+\Rb,
\end{equation}
\begin{equation}
    \Tilde{\Kb}_k^{(i)}=\Pb_{k|k-1}^{(i)}\Hb^{\intercal}[\Sb_{k|k-1}^{(i)}]^{-1}
\end{equation}

Because the number of possibilistic Gaussian components $\bar{N}_k$ used to represent the spatial possibility function grows without bound over time, some techniques are introduced to reduce computational costs, e.g. reducing to a linear increase by exploiting the conditional independence property \cite{eryildirim2016gaussian}, pruning of components with insignificant weights, and merging components those are closely spaced.

\section{Chernoff fusion of  Bernoulli Gaussian max filters}
\label{sec3}


\subsection{General formulation}

The posterior possibility function of a Bernoulli uncertain finite set (UFS)  \cite{ristic2020target} at time $k$ is specified by:
\begin{equation}
f_{k|k}(\Xb|\Zb_{1:k}) = \begin{cases}
q_{k|k}; & \text{ if } X = \emptyset \\
r_{k|k}\,s_{k|k}(\xb); & \text{ if } X = \{\xb\}
 \end{cases}
 \end{equation}

Consider two Bernoulli GMFs processing local sensor measurements: $\Zb_{1:k}^{(1)}$ from sensor 1 and $\Zb_{1:k}^{(2)}$ from sensor $2$.  The corresponding posterior possibility functions are $f^1(\Xb|\Zb_{1:k}^{(1))}$ and $f^2(\Xb|\Zb_{1:k}^{(2)}$, respectively. Then the equation for Chernoff fusion of these two posteriors is as follows:
\begin{equation}
   f^\omega(\Xb|\Zb_{1:k}^{(1)},\Zb_{1:k}^{(2)})=\frac{[f^1(\Xb|\Zb_{1:k}^{(1)})]^{1-\omega}[f^2(\Xb|\Zb_{1:k}^{(2)})]^\omega}{\displaystyle\sup_{\Xb\in\mathcal{F}(\mathcal{X})}\{[f^1(\Xb|\Zb_{1:k}^{(1)})]^{1-\omega}[f^2(\Xb|\Zb_{1:k}^{(2)})]^\omega\}} 
\end{equation}
Let $\Xb=\emptyset$, We can derive the following fusion equations:
\begin{equation}
    q_0^\omega = \frac{(q_0^1)^{1-\omega}(q_0^2)^\omega}{D},
\label{q0}
\end{equation}
where 
\begin{equation}
    D=\max\{(q_0^1)^{1-\omega}(q_0^2)^\omega, (q_1^1)^{1-\omega}(q_1^2)^\omega \sup_{\xb\in\mathcal{X}}[s^1(\xb|\Zb_{1:k}^{(1)})]^{1-\omega}[s^2(\xb|\Zb_{1:k}^{(2)})]^\omega\}
\end{equation}
Let $\Xb=\{\xb\}$, then we will have $f^\omega(\{\xb\})=q_1^\omega s^\omega(\xb)$, and $\sup_{\xb\in\mathcal{X}}f^\omega(\{\xb\})= q_1^\omega$. 
\begin{equation}
    q_1^\omega = \frac{(q_1^1)^{1-\omega}(q_1^2)^\omega\sup_{\xb\in\mathcal{X}}[s^1(\xb|\Zb_{1:k}^{(1)})]^{1-\omega}[s^2(\xb|\Zb_{1:k}^{(2)})]^{\omega}}{D}
    \label{q1}
\end{equation}
\begin{equation}
\begin{split}
    s^\omega(\xb) &= \frac{(q_1^1)^{1-\omega}(q_1^2)^\omega[s^1(\xb|\Zb_{1:k}^{(1)})]^{1-\omega}[s^2(\xb|\Zb_{1:k}^{(2)})]^{\omega}}{q_1^\omega D} \\
    &=\frac{[s^1(\xb|\Zb_{1:k}^{(1)})]^{1-\omega}[s^2(\xb|\Zb_{1:k}^{(2)})]^{\omega}}{\sup_{\xb\in\mathcal{X}}[s^1(\xb|\Zb_{1:k}^{(1)})]^{1-\omega}[s^2(\xb|\Zb_{1:k}^{(2)})]^{\omega}}
    \end{split}
    \label{s}
\end{equation}

Note that in the special case that if sensor measurements are conditionally independent, both $\omega$ and $1-\omega$ will be replaced with $1$, and the fusion equation will become
    \begin{equation}
   f^\omega(\Xb|\Zb_{1:k}^{(1)},\Zb_{1:k}^{(2)})=\frac{[f^1(\Xb|\Zb_{1:k}^{(1)})][f^2(\Xb|\Zb_{1:k}^{(2)})]}{\displaystyle\sup_{\Xb\in\mathcal{F}(\mathcal{X})}\{[f^1(\Xb|\Zb_{1:k}^{(1)})][f^2(\Xb|\Zb_{1:k}^{(2)})]} 
\end{equation}
Let $\Xb = \emptyset$, we will have the following equations:
\begin{equation}
    q_0 = \frac{(q_0^1)(q_0^2)}{K},
\label{q0*}
\end{equation}
where 
\begin{equation}
    K=\max\{q_0^1q_0^2, q_1^1q_1^2\sup_{\xb\in\mathcal{X}}[s^1(\xb|\Zb_{1:k}^{(1)})][s^2(\xb|\Zb_{1:k}^{(2)})]\}
\end{equation}
Let $\Xb=\{\xb\}$, we can derive
\begin{equation}
    q_1 = \frac{q_1^1q_1^2\sup_{\xb\in\mathcal{X}}[s^1(\xb|\Zb_{1:k}^{(1)})][s^2(\xb|\Zb_{1:k}^{(2)})]}{K}
    \label{q1*}
\end{equation}
\begin{equation}
\begin{split}
    s(\xb) &= \frac{q_1^1 q_1^2[s^1(\xb|\Zb_{1:k}^{(1)})][s^2(\xb|\Zb_{1:k}^{(2)})]}{q_1 K} \\
    &=\frac{[s^1(\xb|\Zb_{1:k}^{(1)})][s^2(\xb|\Zb_{1:k}^{(2)})]}{\sup_{\xb\in\mathcal{X}}[s^1(\xb|\Zb_{1:k}^{(1)})][s^2(\xb|\Zb_{1:k}^{(2)})]}
    \end{split}
    \label{s*}
\end{equation}

\subsection{Formulation for Gaussian-max filters}
If $s^1(\xb)$ and $s^2(\xb)$ are represented in Gaussian-max form (time dependence is omitted) as
\begin{equation}
    s^1(\xb)=\textstyle\max_{1\leq i\leq N^{(1)}}w_i^{(1)}\mathcal{N}(\xb;\mb_i^{(1)},\Pb_i^{(1)}),
\end{equation}
and
\begin{equation}
     s^2(\xb)=\textstyle\max_{1\leq j \leq N^{(2)}}w_j^{(2)}\mathcal{N}(\xb;\mb_j^{(2)},\Pb_j^{(2)}).
\end{equation}
where $\mathcal{N}$ is the possibilistic Gaussian function. Then, equations (\ref{q0}), (\ref{q1}) and (\ref{s}) can be written as:
\begin{equation}
      q_0^\omega = \frac{(q_0^1)^{1-\omega}(q_0^2)^\omega}{R},  
\label{q0max}
\end{equation}
\begin{equation}
     q_1^\omega = \frac{(q_1^1)^{1-\omega}(q_1^2)^\omega\alpha}{R},
     \label{q1max}
\end{equation}
and
\begin{equation}
    \begin{split}
     s^\omega(\xb) &= \frac{\max_{1\leq i \leq N^{(1)}}\max_{1\leq j\leq N^{(2)}}w_{ij}^{(1)(2)}\mathcal{N}(\xb;\mb_{ij}^{(1)(2)},\Pb_{ij}^{(1)(2)})}{\alpha} \\
     &=\textstyle\max_{n=1}^{N^*} \tilde{w}_{n}\mathcal{N}(\xb;\mb_{n}^{(1)(2)},\Pb_{n}^{(1)(2)})
     \end{split}
     \label{smax}
\end{equation}
where 
\begin{equation}
    R= \max\biggl\{(q_0^1)^{1-\omega}(q_0^2)^\omega,\left[ (q_1^1)^{1-\omega}(q_1^2)^\omega \alpha \right]\biggr\},
\end{equation}
\begin{equation}
    w_{ij}^{(1)(2)} = (w_i^{(1)})^{1-\omega}(w_j^{(2)})^{\omega}\cdot\mathcal{N}\biggl(\mb_i^{(1)}-\mb_j^{(2)};\mathbf{0}, \frac{P_i^{(1)}}{1-\omega}+\frac{\Pb_j^{(2)}}{\omega}\biggr)
\end{equation}
\begin{equation}
    \Pb_{ij}^{(1)(2)}=[(1-\omega)(\Pb_i^{(1)})^{-1}+\omega(\Pb_j^{(2)})^{(-1)}]^{-1}
\end{equation}
\begin{equation}
    \mb_{ij}^{(1)(2)} = \Pb_{ij}^{(1)(2)}\left[ (1-\omega)(\Pb_i^{(1)})^{-1}\mb_i^{(1)}+\omega(\Pb_j^{(2)})^{-1}\mb_j^{(2)}\right]
\end{equation}
\begin{equation}
    \tilde{w}_n=\frac{w_{n}}{\alpha}
\end{equation}
\begin{equation}
        n= (j-1)*N^{(1)}+i, \; i=1,\cdots,N^{(1)} \; \text{and} \; j=1,\cdots, N^{(2)} 
\end{equation}
\begin{equation}
     N^*= N^{(1)}*N^{(2)}
\end{equation}
\begin{equation}
    \alpha=\textstyle\max_{1\leq i \leq N^{(1)}}\max_{1\leq j \leq N^{(2)}}w_{ij}^{(1)(2)}
\end{equation}
In the Gaussian-max form, for the special case of independent sensors, similarly we will rewrite \ref{q0max}, \ref{q1max}, \ref{smax} as the following equations:
\begin{equation}
      q_0 = \frac{q_0^1q_0^2}{G},  
\end{equation}

\begin{equation}
     q_1= \frac{q_1^1q_1^2\alpha ^*}{G},
\end{equation}
and
\begin{equation}
    \begin{split}
     s(\xb) &= \frac{\max_{1\leq i \leq N^{(1)}}\max_{1\leq j\leq N^{(2)}}w_{ij}^{(1)(2)}\mathcal{N}(\xb;\mb_{ij}^{(1)(2)},\Pb_{ij}^{(1)(2)})}{\alpha ^*} \\
     &=\textstyle\max_{1\leq n \leq N^*} \tilde{w}_{n}\mathcal{N}(\xb;\mb_{n}^{(1)(2)},\Pb_{n}^{(1)(2)})
     \end{split}
\end{equation}
Where
\begin{equation}
    G= \max\biggl\{q_0^1q_0^2,q_1^1q_1^2 \alpha ^*\biggr\},
\end{equation}
\begin{equation}
    w_{ij}^{(1)(2)} = w_i^{(1)}w_j^{(2)}\cdot\mathcal{N}\biggl(\mb_i^{(1)}-\mb_j^{(2)};\mathbf{0}, P_i^{(1)}+\Pb_j^{(2)}\biggr)
\end{equation}
\begin{equation}
    \Pb_{ij}^{(1)(2)}=[(\Pb_i^{(1)})^{-1}+(\Pb_j^{(2)})^{-1}]^{-1}
\end{equation}
\begin{equation}
    \mb_{ij}^{(1)(2)} = \Pb_{ij}^{(1)(2)}\left[(\Pb_i^{(1)})^{-1}\mb_i^{(1)}+(\Pb_j^{(2)})^{-1}\mb_j^{(2)}\right]
\end{equation}
\begin{equation}
    \alpha ^*=\textstyle\max_{1\leq i \leq N^{(1)}}\max_{1\leq j \leq N^{(2)}}w_{ij}^{(1)(2)}
\end{equation}
\begin{equation}
    \tilde{w}_n=\frac{w_{n}}{\alpha ^*}
\end{equation}

Recall that the derivation of Chernoff fusion of Bernoulli Gaussian sum filters in the probabilistic framework  \cite{guldogan2014consensus}, involved an approximation. Note that derivation in the possibilistic framework, presented above, results in the exact solution  (no approximation involved).  

\section{Numerical results}
\label{sec4}
\begin{figure}[htb]
  \centering
  \includegraphics[width=0.7\linewidth]{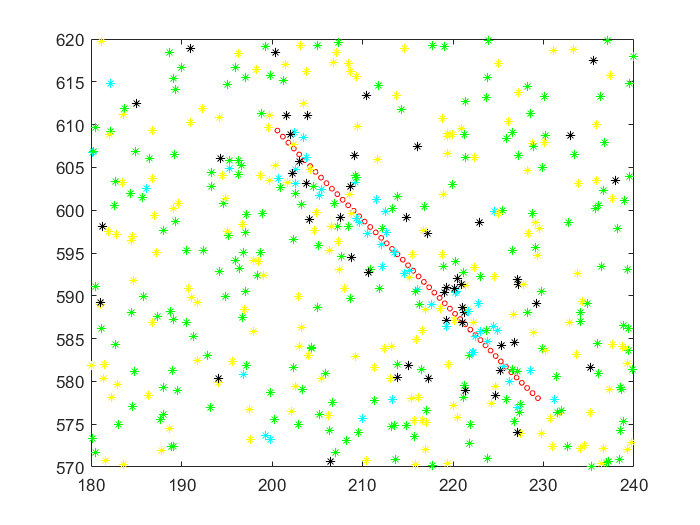}
  \caption{Simulation test setup}
\label{fig1:test_scenario}
\end{figure}

The simulation tests will consider single-target tracking using two sensors in a $60\times60$ km $2$-Dimensional surveillance area. The top-down view of the scenario is shown in Fig. \ref{fig1:test_scenario}, where colored asterisks represent the measurements received by the two sensors, and red circles indicate the true locations of the moving target over time. The state of the target is specified in Cartesian coordinates as $\xb=\left[x, \; \dot x, \; y, \; \dot y\right]^\intercal$. The total tracking time steps is set as $N=50$. A single target with the initial state $\xb = \left[10, 0.3, 55, -0.35\right]^\intercal$ enters in the surveillance area at time $k=1$ and leaves at $k=50$. 

\subsection{Data generation and filter parameters settings}
Sensors are providing standard point measurements. 
The true probability of detection of Sensor $1$ is set to $PD^{(1)}_{\text{\tiny {true}}} = 0.8$ and the true probability of detection of Sensor $2$ is set to $PD^{(2)}_{\text{\tiny {true}}} = 0.6$. The false detections are assumed to be uniformly distributed over the surveillance area with average count $\lambda = 4$ per time step and the measurement noise variances are set as $\sigma_v =2$ for both sensors.

Target motion model is specified by the transitional density (\ref{e:trans}), 
where $\Fb$ and $\Qb$ are 
\begin{equation}
    \Fb = \left[\begin{matrix} 1 & T & 0 & 0 \\ 0 & 1 & 0 & 0 \\ 0 & 0 & 1 & T \\ 0 & 0 & 0& 1 \end{matrix}\right],
    \label{F}
\end{equation}
\begin{equation}
    \Qb = \left[\begin{matrix} \frac{T^3}{3} & \frac{T^2}{2} & 0 & 0 \\ \frac{T^2}{2} & T & 0 & 0 \\ 0 & 0 & \frac{T^3}{3} & \frac{T^2}{2}  \\ 0 & 0 & \frac{T^2}{2} & T  \end{matrix}\right]\mathcal{q},
    \label{Q}
\end{equation}
where $T$ is the sampling interval, and set to $T=2$s, and $\mathcal{q}$ is the level of power spectral density of the corresponding continuous process noise and set to $\mathcal{q}=0.00001$. The probability of target birth is $p_b =0.05$ and probability of target survival is  $p_s=0.99$, while the birth density is measurement driven.   

The transitional possibility matrix is set as:
\begin{equation}
    \Phi = \left[\begin{matrix} 1 & 0.01 \\ 0.01 & 1\end{matrix}\right].
\end{equation}
The Gaussian max filter can operate with partially know probability of detection. Suppose we only know for both sensors that the true probability of detection is in the interval $[0.5,1]$.
 Then, the possibility of detection can be set to $d^1 = 1$ and the possibility of non-detection to $d^0 = 0.5$. These values are obtained by transforming the partially known probability to possibilities \cite{chen2021observer,chen2023bernoulli}.
 
\subsection{Independent sensors}
\begin{figure}[htb]
  \centering
  \includegraphics[width=\linewidth]{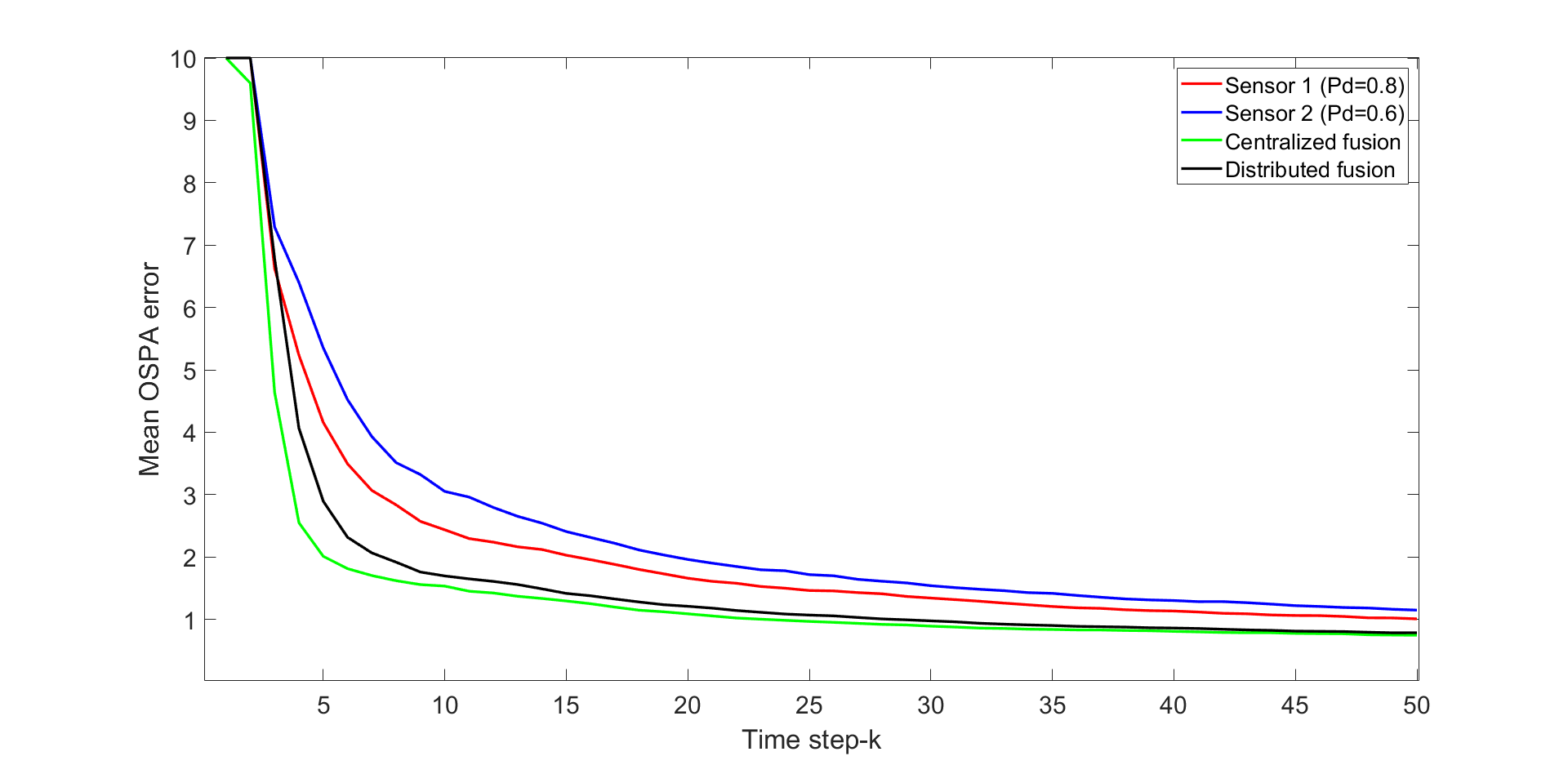}
  \caption{Fusion with independent sensors}
\label{fig2:independent}
\end{figure}

Firstly, we will consider the case of independent sensors, and evaluate the tracking performance using The Optimal Sub-pattern Assignment (OSPA) metrics. We consider Bernoulli Gaussian-max filters from two individual sensors, the centralized fusion, and distributed Chernoff fusion (derived above). Fig. \ref{fig2:independent} shows the averaging OSPA results for $2000$ Monte Carlo simulation runs over time.
We can see both the centralized fusion and distributed fusion show a better performance than Bernoulli Gaussian-max filters of y individual sensors. Distributed Chernoff fusion is cautios and therefore somewhat worse than  the centralized fusion.

\subsection{Unknown dependence between the sensors}

\begin{figure}[htb]
  \centering
  \includegraphics[width=\linewidth]{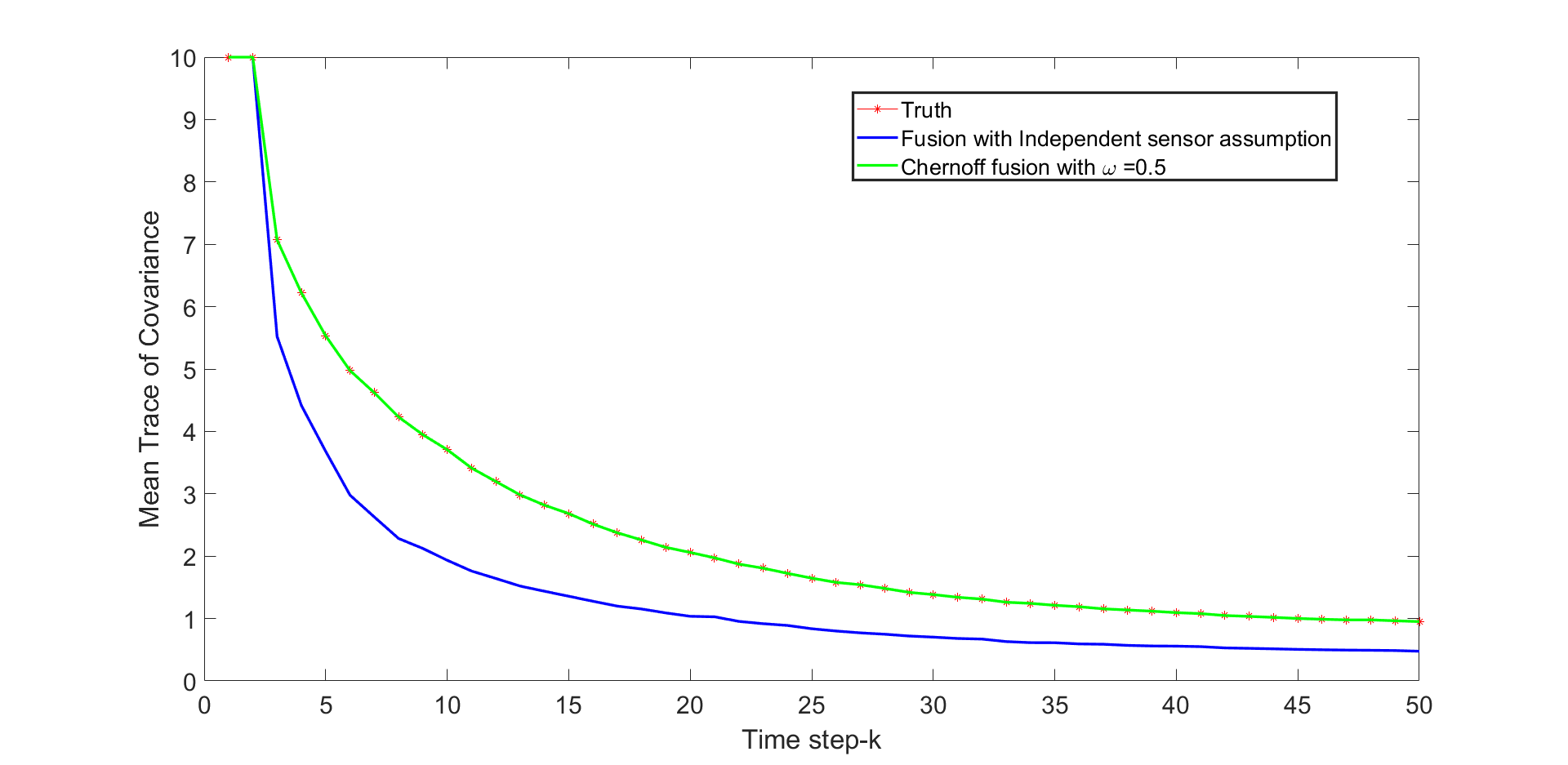}
  \caption{Mean covariance of fused estimates in $1000$ Monte Carlo runs}
\label{fig3:chernoff}
\end{figure}

Next we consider the case of totally dependent sensors: both Bernoulli-GMF-1 and Bernoulli-GMF-2 obtain measurements from the same sensor. We compare the centralised fusion and distributed Chernoff fusion of two Bernoulli filters. The performance measure this time is the size of the fused covariance matrix (measured by its trace). 
Fig. \ref{fig3:chernoff} shows that Chernoff results in  correct covariance for fused estimates,  while the centralised is overoptimistic (because it assumes independence of sources of data). 

\section{Conclusions}
\label{sec5}

In this paper, we formulate close form analytical formulation of Chernoff fusion for Bernoulli Gaussian max filters, in the framework of possibility theory. The derivation is exact, without any approximations (unlike the probabilistic case). Simulation results demonstrate that the proposed Chernoff fusion algorithms in the possibilistic framework, works in both scenarios with and without knowledge of sensor dependence, and with only partially known probability of detection.

\section*{Acknowledgment}
This research is supported by an Australian Government Research Training Program (RTP) Scholarship.

\bibliographystyle{elsarticle-num}
\bibliography{main}

\begin{thebibliography}{10}
\expandafter\ifx\csname url\endcsname\relax
  \def\url#1{\texttt{#1}}\fi
\expandafter\ifx\csname urlprefix\endcsname\relax\def\urlprefix{URL }\fi
\expandafter\ifx\csname href\endcsname\relax
  \def\href#1#2{#2} \def\path#1{#1}\fi

\bibitem{mahler2007statistical}
R.~Mahler, Statistical multisource-multitarget information fusion, Artech, 2007.

\bibitem{ristic2013bernoulli}
B.~Ristic, J.~Sherrah, Bernoulli filter for joint detection and tracking of an extended object in clutter, IET Radar, Sonar \& Navigation 7~(1) (2013) 26--35.

\bibitem{vo2012multi}
B.~T. Vo, C.~M. See, N.~Ma, W.~T. Ng, Multi-sensor joint detection and tracking with the bernoulli filter, IEEE Transactions on Aerospace and Electronic Systems 48~(2) (2012) 1385--1402.

\bibitem{ristic2013tutorial}
B.~Ristic, B.-T. Vo, B.-N. Vo, A.~Farina, A tutorial on bernoulli filters: theory, implementation and applications, IEEE Transactions on Signal Processing 61~(13) (2013) 3406--3430.

\bibitem{julier2006empirical}
S.~J. Julier, An empirical study into the use of chernoff information for robust, distributed fusion of gaussian mixture models, in: 2006 9th International Conference on Information Fusion, IEEE, 2006, pp. 1--8.

\bibitem{guldogan2014consensus}
M.~B. Guldogan, Consensus bernoulli filter for distributed detection and tracking using multi-static doppler shifts, IEEE Signal Processing Letters 21~(6) (2014) 672--676.

\bibitem{gunay2014approximate}
M.~G{\"u}nay, U.~Orguner, M.~Demirekler, Approximate chernoff fusion of gaussian mixtures using sigma-points, in: 17th International Conference on Information Fusion (FUSION), IEEE, 2014, pp. 1--8.

\bibitem{houssineau2018detection}
J.~Houssineau, Detection and estimation of partially-observed dynamical systems: an outer-measure approach, arXiv preprint arXiv:1801.00571 (2018).

\bibitem{ristic2020target}
B.~Ristic, J.~Houssineau, S.~Arulampalam, Target tracking in the framework of possibility theory: The possibilistic bernoulli filter, Information Fusion 62 (2020) 81--88.

\bibitem{cai2022possibility}
H.~Cai, J.~Houssineau, B.~A. Jones, M.~Jah, J.~Zhang, Possibility generalized labeled multi-bernoulli filter for multitarget tracking under epistemic uncertainty, IEEE Transactions on Aerospace and Electronic Systems 59~(2) (2022) 1312--1326.

\bibitem{chen2023bernoulli}
Z.~Chen, B.~Ristic, D.~Y. Kim, Bernoulli filter for extended target tracking in the framework of possibility theory, IEEE Transactions on Aerospace and Electronic Systems 59~(6) (2023) 9733--9739.

\bibitem{houssineau2022decentralised}
J.~Houssineau, H.~Cai, M.~Uney, E.~Delande, Decentralised possibilistic inference with applications to target tracking, arXiv preprint arXiv:2209.12245 (2022).

\bibitem{bishop2018spatio}
A.~N. Bishop, J.~Houssineau, D.~Angley, B.~Ristic, Spatio-temporal tracking from natural language statements using outer probability theory, Information Sciences 463 (2018) 56--74.

\bibitem{jeremie_linear_21}
J.~Houssineau, A linear algorithm for multi-target tracking in the context of possibility theory, IEEE Transactions on Signal Processing 69 (2021) 2740--2751.

\bibitem{eryildirim2016gaussian}
A.~Eryildirim, M.~B. Guldogan, A gaussian mixture bernoulli filter for extended target tracking with application to an ultra-wideband localization system, Digital Signal Processing 57 (2016) 1--12.

\bibitem{chen2021observer}
Z.~Chen, B.~Ristic, J.~Houssineau, D.~Y. Kim, Observer control for bearings-only tracking using possibility functions, Automatica 133 (2021) 109888.

\end{thebibliography}
\end{document}